\documentclass[conference]{IEEEtran}
\usepackage[utf8]{inputenc}
\usepackage{graphicx}
\usepackage{algorithm,algpseudocode}
\usepackage{array}
\usepackage{amsmath}
\usepackage{amssymb}
\usepackage{amsthm}
\usepackage{mdwmath}
\usepackage{diagbox}
\usepackage{flexisym}
\usepackage{color,soul}
\usepackage{textcomp}
\usepackage{subfig}
\usepackage{multirow}
\usepackage{multicol}
\usepackage[export]{adjustbox}
\usepackage[english]{babel}
\usepackage[utf8]{inputenc}
\usepackage{comment}
\usepackage[binary-units]{siunitx}
\usepackage{makecell}
\usepackage{pgfplots,filecontents}
\usepgfplotslibrary{groupplots}
\usepackage{hyperref}
\usepackage{filecontents}
\usepackage{footnote}
\usepackage{wasysym}
\usepackage{color,soul}
\usepackage{xcolor}
\usepackage{lipsum}
\usepackage{wrapfig}
\setcellgapes{1pt}
\usepackage[T1]{fontenc}
\usepackage{siunitx}
\usepackage{footmisc}
\pgfplotsset{compat=newest}
\pgfplotsset{%
   every mark/.append style={scale=1.0},
   width=6.1cm,compat=1.3
}

\newcolumntype{P}[1]{>{\centering\arraybackslash}p{#1}}
\newcolumntype{M}[1]{>{\centering\arraybackslash}m{#1}}

\newcommand{\Data}{\mathit{Data}}
\newcommand{\pred}{\mathit{pred}}
\newcommand{\suc}{\mathit{succ}}
\newcommand{\sched}{\mathit{plc}}
\newcommand{\CPU}{\mathit{CPU}}
\newcommand{\MEM}{\mathit{MEM}}
\newcommand{\STOR}{\mathit{STOR}}
\newcommand{\BW}{\mathit{BW}}

\begin{document}
\title{MAPO: A Multi-Objective Model for IoT Application Placement in a Fog Environment}

\author{\IEEEauthorblockN{
{Narges Mehran},
{Dragi Kimovski},
{Radu Prodan}}
\IEEEauthorblockN{Institute of Information Technology, Alpen-Adria-Universit{\"a}t Klagenfurt, Austria\\
Email: \{name\}.\{surname\}@aau.at}
}
\maketitle

\begin{abstract}
The emergence of the Fog computing paradigm that leverages in-network virtualized resources raises important challenges in terms of resource and IoT application management
in a heterogeneous environment offering only limited computing resources.
In this work, we propose a novel Pareto-based approach for application  placement close  to  the data sources called Multi-objective IoT Application Placement  in  fOg (MAPO). MAPO models applications based on a finite state machine and uses three conflicting optimization objectives, namely completion time, energy consumption, and economic cost, considering both the computation and communication aspects. In contrast to existing solutions that optimize a single objective value, MAPO enables multi-objective energy and cost-aware application placement. To  evaluate  the  quality of the MAPO  placements, we created both simulated and real-world testbeds tailored for a set of medical IoT application case studies.
Compared to the state-of-the-art approaches, MAPO reduces the economic cost by up to 27\%, while decreasing the energy requirements by 23\texttt{-}68\%, and optimizes the completion time by up to 7.3 times. 

\end{abstract}


\section{Introduction} \label{Introduction}
It is predicted that the number of connected \emph{Internet of Things (IoT)} devices will increase by 2020 to more than 20 billions, requiring appropriate computational and storage capabilities offered by the Cloud Data Centers (CDCs)\footnote{http://www.gartner.com/newsroom/id/3165317}. However, the latency to reach the current CDCs for many large-scale systems, such as IoT based medical applications, can be unacceptably high. It is therefore essential to explore the possibility of executing such applications in 
lower Cloud-Fog hierarchy
physically closer to the IoT devices. As the utilization of in-network, Edge \cite{Nardelli2017Schahram} and micro-data centers \cite{Aazam2015} can improve the performance of the IoT systems \cite{Varghese2016}, \emph{Fog computing} \cite{bonomi2012fog} emerged as a new paradigm that partially moves the processing of latency-sensitive IoT applications from the Cloud to the Edge of the networks, where the data is generated \cite{Hajibaba2014}.\par

The emergence of the Fog paradigm as a computing environment for the IoT systems raises several challenges in terms of resource and application management \cite{kimovski2018adaptive}. Therefore, it becomes essential to efficiently manage the execution of distributed applications in a highly heterogeneous environment with limited computing capacities. However, only a few state-of-the-art research works explored this problem so far \cite{Aazam2015,Sun2018} by optimizing either a single constrained objective or a set of weighted objectives, and omitting important parameters such as energy efficiency and resource utilization.\par
We explore in this work a novel approach called \emph{Multi-objective IoT Application Placement in fOg (MAPO)}, tailored to specific IoT application modelled as a set of lightweight interconnected components, implemented either as micro-services or as functional modules \cite{alam2018orchestration}, in a heterogeneous Fog computing environment.
To tackle this problem, we apply a genetic multi-objective optimization algorithm that considers three conflicting criteria (i.e., completion time, energy consumption and economic cost) to approximate the Pareto set of optimized placements of the application components on the available Fog devices. On top of it, we implement a decision making strategy for selecting a single placement solution from the Pareto set based on the applications requirements.\par
Unlike existing solutions that optimize a single constrained objective or a set of weighted objectives, our approach delivers lower execution time, reduced power consumption, and provides cost-aware application management.
To evaluate the performance of the Pareto placement method, we create elaborate scenarios in both simulated and real-world Fog environments, specifically tailored for a set of medical IoT application case studies. We compare MAPO with two related state-of-the-art methods \cite{Skarlat2017QoS-aware,gupta2017ifogsim} and show its potential to reduce the application completion time (per request) by up to seven times and to decrease the energy requirements by 68\%.\par
The paper is organized as follows. Section \ref{RelatedWork} provides a survey of the relevant related work. Section \ref{Model} elaborates the model underneath our approach, followed by the architecture for the placement strategy in Section \ref{Overview}. Section \ref{CaseStudies} describes the case studies along with their application models. Section \ref{simulation} defines the simulation experimental setup and its results, followed by the real-world testbed results in Section \ref{testbed}. Finally, Section \ref{Conclusion} concludes the paper.\par

\section{Related work} \label{RelatedWork}
Resource provisioning and application placement are essential problems in Fog computing for the next generation IoT applications. 

The authors in \cite{Aazam2015} proposed a resource management model and a framework for Fog environments that assumes a heterogeneous pool of IoT devices with an unpredictable utilization rate. The resource provisioning considers the behaviour of the costumers in relation to the service type and price, and aims to provide mechanisms for reducing the execution cost in Fog, thus encouraging the costumers to use this environment.\par

The authors in \cite{Sun2018} investigated a two-level scheduling model that divides the Fog devices into multiple geographical clusters. The resource scheduling is performed among various Fog clusters and internally within a given cluster. In the latter method, a weighted multi-objective optimization technique reduces the service latency and improves the overall stability of task executions. \par 



The work in \cite{Pham2016} proposed a heuristic algorithm for task scheduling in a Cloud-Fog system, allowing providers to utilize their own Fog nodes together with rented Cloud resources, and providing aggregated cost and makespan tradeoffs for offloading large-scale applications. 
Furthermore, the authors proposed in \cite{gupta2017ifogsim} a placement strategy in a Cloud-Fog environment considering a specific Fog device or CDC for computing the last component of every application. Then, other components (predecessors of the last component) of that application are placed on the available lower computational devices in the Cloud-Fog hierarchy.

The authors studied in \cite{Bittencourt2017} the impact of three scheduling methods (i.e. concurrent, first-come-first-served and delay priority) on the quality of service (QoS) in a Fog environment using two applications types: a video surveillance and object tracking as a delay tolerant application and an electroencephalography tractor beam game as a near real-time application. The results revealed that the concurrent scheduling method produced a resource contention with a high delay in both cases, and a lower network use with reduced communication between the Fog and the Cloud.\par 

The authors in \cite{xia2018combining} proposed four combined placement heuristics such as ``Fog nodes ordering'' and ``components ordering'' optimizing the weighted average latency for a large set of heterogeneous applications, constrained by a set of functional parameters, including computing and storage resources.\par

The authors in \cite{Skarlat2017QoS-aware} proposed an integer programming model for optimizing the placement between the IoT applications and Fog resources considering the execution cost, which requires periodical updates for guaranteeing QoS parameters.\par 

All of these works focused on optimizing either a single constrained objective, or a set of weighted objectives combined in a linear function. MAPO improves over these approaches by representing the application placement as a multi-objective optimization problem that takes into account user and provider-centred conflicting criteria. Furthermore, MAPO implements a Pareto-based decision making strategy to guarantee selection of placement solutions according to the applications requirements.

\section{Model} \label{Model}
This section presents a formal model and a set of essential definitions important for this work.

\subsection{Application model}  \label{App_Model}
We represent an IoT application $A = \left(M, \Sigma, \Gamma, m_1, Data_1, \eta\right)$ as a state machine consisting of six parts: a set of $x$ lightweight interconnected \emph{components} or states $M = \bigcup_{i=1}^x m_i$; an alphabet $\Sigma$ representing all possible \emph{data items} generated by all components $m_i \in M$; a \emph{transition function} $\Gamma: M \times \Sigma \to M \times \Sigma$, where $\Gamma\left(m_i, Data_i\right) = \left(m_j, Data_j\right)$ represents the outcome of processing the data item $Data_i$ using component $m_i$ and producing a new data item $Data_j$ to be transferred and processed by the successor $m_j$; one \emph{start component} or state $m_1 \in M$ 
receiving the input data $Data_1$ from the IoT devices and triggering the transition $\Gamma\left(m_1, Data_1\right)$; and a set of \emph{accepting components} $\eta \subset M$.

\subsection{Resource model}  \label{Res_Model}
We consider a set of $y$ heterogeneous Fog \emph{devices} $RS=\bigcup_{j=1}^y r_j$, where an individual device $r_{j}$ = ($\CPU_j$, $\MEM_j$, $\STOR_j$) defines the number of instructions per second $\CPU_j$, the memory size $\MEM_j$, and the permanent storage size $\STOR_j$ available \cite{Mahmud2018}.
The components are placed (or deployed) as services running in virtualized container environments \cite{pahl2017cloud}. 
Proper execution of a component $m_{i}$ requires a minimal amount of resources, defined in terms of the number of instructions $INSTR\left(m_i\right)$, processing speed $\CPU\left(m_i\right)$, memory $\MEM\left(m_i\right)$, and storage $\STOR\left(m_i\right)$ requirements.

We define the \emph{placement} of an IoT application $A$ 
on a set of Fog devices $R\ \subset\ RS$ as a function $\sched: A \to R$ that maps each component $m_i \in A$ on a device $r_j \in R$.
The image of the placement function $R = \sched\left(A\right)$ is the set of devices where the application runs.

\subsection{Execution model}\label{Exe_Model}
Users of IoT applications submit them to MAPO for placement on the Fog infrastructure and trigger their executions. Every application instance has a sequential execution thread. Within one execution, every component instance $m_i$ has one \emph{successor} $\suc\left(m_i\right) =  m_j$, where $\Gamma\left(m_i, Data_i\right) = \left(m_j, Data_j\right)$ and $Data_j$ represents the output of $m_i$ transferred as input to $m_j$ for subsequent processing.
Similarly, every component instance $m_i$ has one \emph{predecessor} $\pred\left(m_j\right) = m_i$ that provides its output data $Data_j$ as input for processing, where $ \Gamma\left(m_j, Data_j\right) = \left(m_i, Data_i\right)$.
The starting component $m_1$ has no predecessor ($\pred\left(m_1\right) = \emptyset$), and the accepting component (in one execution) $m_x \in \eta$ has no successor ($\suc\left(m_x\right) = \emptyset$).

\subsection{Optimization objectives} \label{OO}
We consider three objectives for placing an application on the Fog.

\subsubsection{Completion time} $T\left(m_i,r_j\right)$ of a component $m_i$ on a device $r_j = \sched\left(m_i\right)$ is the sum between the completion time of its predecessor $\pred\left(m_i\right)$ on device $r_k = \sched\left(\pred\left(m_i\right)\right)$, the time for receiving its input data $Data_i$, and its computation time $t\left(m_i, r_j\right)$: 
\begin{equation}
    T\left(m_i,r_j\right)= \left \{
  	\begin{array}{@{ }l@{ }l@{ }}
    	t\left(m_i, r_j\right), & \pred(m_i) = \emptyset;\\
    	T\left(\pred\left(m_i\right),r_k\right) +
		\frac{Data_i}{\BW_{k,j}} + t\left(m_i, r_j\right), & \pred\left(m_i\right) \neq \emptyset,
    \end{array}
	\right.
    \label{eq1}
\end{equation}
where the \emph{computation time} $t\left(m_i,r_j\right)$ of a component $m_i$ on a Fog device $r_j$ is the ratio between the component workload defined by its number of instructions $INSTR\left(m_i\right)$ and the processing speed $\CPU_j$ or resource $r_j$ \cite{Vindm2018}:
\begin{equation}
	t\left(m_i, r_j\right) = \frac{INSTR\left(m_i\right)}{\CPU_j},
    \label{eq2}
\end{equation}
and $BW_{k,j}$ is network bandwidth between devices $r_k$ and $r_j$.

The completion time of an application $A$ placed on the set $R = \sched\left(A\right)$ of Fog devices is (where $\suc\left(m_x\right) = \emptyset$):
\begin{equation}
T\left(A,R\right) = T\left(m_x, \sched\left(m_x\right)\right).
\label{eq3}
\end{equation}


\subsubsection{Energy consumption} $E\left(m_i,r_j\right)$ of a component $m_i$ executed on a device $r_j$ is the sum of the computation $E_{p}\left(m_i,r_j\right)$, communication $E_{m}\left(r_k,r_j\right)$ to retrieve data from the predecessor placed on device $r_k = \sched\left(\pred\left(m_i\right)\right)$, and the \emph{static energy} $E_{s}\left(m_i,r_j\right)$ for maintaining the device active:
\begin{equation}
E\left(m_i,r_j\right) = E_p\left(m_i,r_j\right) +E_m\left(r_k,r_j\right) + E_s\left(m_i,r_j\right).
\label{eq5}
\end{equation}

The \emph{computation energy} consumed for executing a single component $m_i$ on a device $r_j$ is:
\begin{equation}
E_{p}\left(m_i,r_j\right) = \varrho_j^p \cdot t\left(m_i,r_j\right),
\label{eq6}
\end{equation}
where $\varrho_j^p$ is the computational power consumption of $r_j$.

The \emph{communication energy} required by a network interface of device $r_j$ to receive and process a data of size $Data_i$ from another device $r_k$ (including all the switching equipment, and the radio communication subsystem) is:
\begin{equation}
\begin{aligned}
E_{m}\left(r_k,r_j\right) = {\varrho_j^m}\cdot{\frac{Data_i}{BW_{k,j}}} + \epsilon_j,
\end{aligned}
\label{eq7}
\end{equation}
where $\varrho_j^m$ is the power consumption of $r_j$ for receiving a data item and $\epsilon_{j}$ is a hardware-related constant \cite{Vindm2018}.

The energy $E\left(A,R\right)$ of executing an application $A$ is the total energy consumed by its components:
\begin{equation}
E\left(A,R\right) = \sum\limits_{\forall m_i \in A\ \land\ \sched\left(m_i\right)=r_{j}}E\left(m_i,r_j\right).
\label{eq8}
\end{equation}

\subsubsection{Economic cost} $C\left(m_i,r_j\right)$ of executing a component $m_i$ on $r_j$ is the sum of its processing, storage, and communication costs:
\begin{equation}
C\left(m_i,r_j\right) = t\left(m_i,r_j\right) \cdot CP_j + Data_i \cdot CS_j + \frac{Data_i}{\BW_{k,j}} \cdot CR_j, 
\label{eq9}
\end{equation}
where $CP_j$, $CS_j$ and $CR_j$ are the processing, storage and ingress communication costs of device $r_j$ per second.

The cost $C(A,R)$ of executing an application on a set $R$ of Fog devices is the sum of the cost of executing its components:
\begin{equation}
C\left(A,R\right) = \sum\limits_{\forall m_i \in A\ \land\ \sched\left(m_i\right)=r_{j}} C\left(m_i,r_j\right).
\label{eq10}
\end{equation}

\subsection{Problem definition}
A multi-objective optimization identifies one or more tradeoff solutions that optimize predefined and possibly conflicting objective functions within a given search space.
Multi-objective optimization problems are typically NP-complete considering the completion time
as one of the $o \geq 2$ objective functions $f_1\left(\vec{\gamma}\right), f_2\left(\vec{\gamma}\right), \ldots, f_o\left(\vec{\gamma}\right)$ for which the global maximum or minimum needs to be identified. In these functions, $\vec{\gamma}=\left(\gamma_1, \gamma_2, \ldots, \gamma_\kappa\right)$ represents a set of so-called decision variables within a search space $Y$, where $\vec{\gamma} \in Y$ and $\kappa$ is the number of space dimensions. A given solution $\vec{\alpha} \in Y$ \emph{dominates} another solution $\vec{\beta} \in Y$ only if it is better in respect to all objectives: $f_u\left(\vec{\alpha}\right) \leq f_u\left(\vec{\beta}\right), \forall u \in [1,o]$, and $\exists v \in [1,o]$ such that $f_v\left(\vec{\alpha}\right) < f_v\left(\vec{\beta}\right)$. The resulting set of non-dominated solutions is called \emph{Pareto optimal set} in the search space $Y$ and represents a set of tradeoff values among the objective functions. The Pareto optimal set forms the \emph{Pareto frontier} of finite points of tradeoff solutions.

For our problem, we consider three optimization objectives ($o=3$), completion time, energy consumption, and economic cost, described in Section~\ref{OO}. The set of decision variables is the set of application components to be placed onto the Fog devices: $\vec{\gamma} = R$.
Each decision variable $\gamma_i$ is the placement of one component $m_i$ onto a Fog device: $\gamma_i = \sched\left(m_i\right)$. Our goal for an application $A$ is to find a placement $\sched(A)$ that assigns all its components to the set $R$ of Fog devices that minimizes the three considered objectives and fulfills their resource requirements:
\begin{equation}
\left\{
\arraycolsep=0pt
\begin{array}{ll}
f_1(A, R) = \min\limits_{\sched(A) = R} T\left(A,R\right);\\
f_2(A, R) = \min\limits_{\sched(A) = R} E\left(A,R\right);\\
f_3(A, R) = \min\limits_{\sched(A) = R} C\left(A,R\right);\\
\forall m_i \in A, r_j = \sched\left(m_i\right), & \CPU\left(m_i\right) <  \CPU_j \land\\
& \MEM\left(m_i\right) < \MEM_j\land\\
& \Data_i < \STOR_j.
\end{array}
\right.
\label{eq11}
\end{equation}

\section{Architecture overview} \label{Overview}

\begin{figure}
	\centering
    \captionsetup{justification=centering}
    \includegraphics[width=0.8\columnwidth]{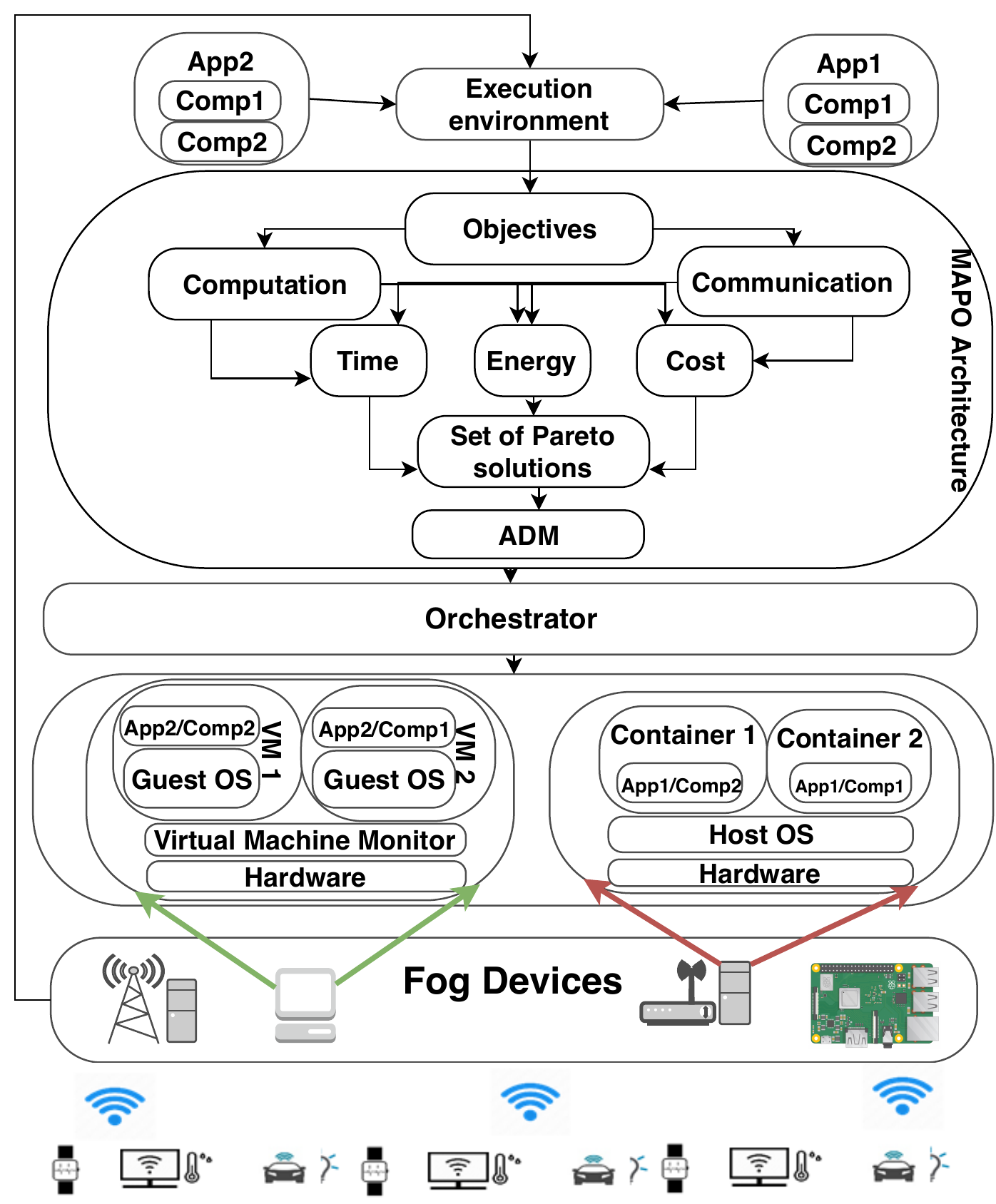}
	\caption{MAPO overall architecture design.}
    \label{figArch}
\end{figure}
Figure~\ref{figArch} describes MAPO architecture through specific case study, in which various application owners deploy their applications composed of multiple inter-connected components.
The applications are deployed in the Fog through the MAPO execution environment, which acts as an interface to the MAPO architecture.
Afterwards, it identifies the inter-connections among the application components following the transition function, and provides the placement constraints to the multi-objective optimization algorithm that searches for a Pareto set of tradeoff solutions considering the objectives described in Section \ref{Model}. The multi-objective optimization follows a relatively fast evolutionary Non-Dominated Sorting Genetic Algorithm (NSGA-II) \cite{Deb2002} that searches for a set of non-dominated Pareto placements of the application components on a set of Fog devices. For this, it ranks the population according to a fast non-dominated sorting method to prepare elitism and good convergence near the true Pareto optimal set. An automated decision making module selects an appropriate placement based on a low latency strategy, which extends on a simple and computationally efficient a-priori method \cite{kimovski2017multi} assuming that all solutions on the Pareto frontier belong to a single cluster. 
Finally, MAPO delivers the selected placement to an orchestrator built on top of Docker Swarm \cite{hoque2017towards} that operates as an execution controller that 
instantiates the containerized application components on the selected devices as defined in Section \ref{Exe_Model}.

\section{Application case studies} \label{CaseStudies}
\begin{wraptable}{R}{0.4\columnwidth}
\centering
\small
\caption{Application resource requirements.}
\label{tbl:apps}
\begin{tabular}{|@{}c@{}|@{}c@{}|@{}c@{}|@{}c@{}|@{}c@{}|}
\hline
\multirow{2}*{\emph{Application}} & \emph{CPU} & \emph{MEM} & \emph{Storage}\\ 
 & [MI] & \SI{}{[\mega\byte]} & \SI{}{[\mega\byte]}\\ 
\hline
Augmented& 100-- & 10-- & 256--\\ 
reality & 2000 &30 & 512\\
\hline
Insulin & 200 -- & 10 -- & 256 --\\ 
pump & 2000 & 60 & 1024\\ 
\hline
Mental & 200 -- & 10 -- & 256 --\\ 
health care & 2000 & 50 & 512\\ 
\hline
\end{tabular}
\end{wraptable}

\begin{figure}[t]
    \centering
    \includegraphics[width=0.8\columnwidth]{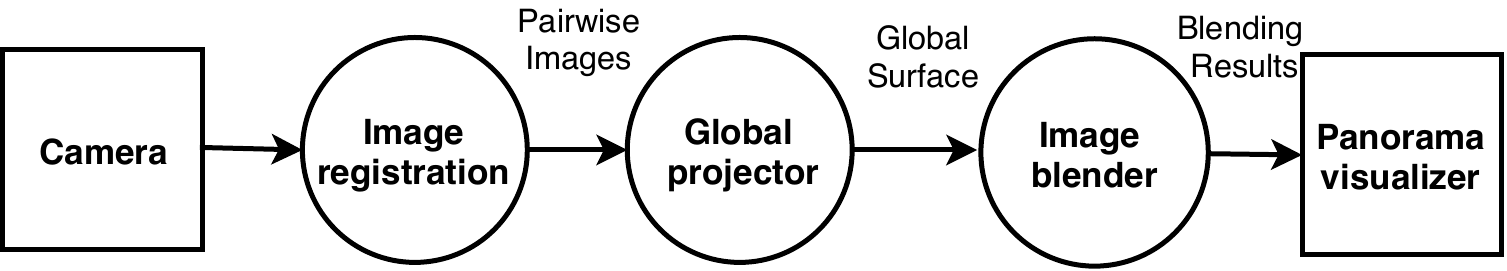}
	\caption{Augmented reality application.}
    \label{figAR}
\end{figure}

We selected three application case studies from the field of next generation medical applications with different computing and storage demands, summarized in Table~\ref{tbl:apps}. 
We modeled every application as a state machine and labelled its transitions with the data items transferred between components, as presented in Section \ref{App_Model}.
\paragraph{Augmented reality}
The medical augmented reality application broadens the surgeon's perception of a surgical scene by merging the isolated information of the live endoscope \cite{bergen2016stitching}. The utilization of the Fog can significantly reduce the processing and communication latency, helping the medical professionals in close to real time scenarios. The application, divided in multiple components shown in Figure~\ref{figAR}, fetches streamed endoscopic images\footnote{\url{http://visurge.wp.itec.aau.at/}} at specific rates (e.g., between 15 and 30 frames per second) and stitches them into a single large panorama view.

\begin{figure}[t]
	\centering
    \includegraphics[width=0.8\columnwidth]{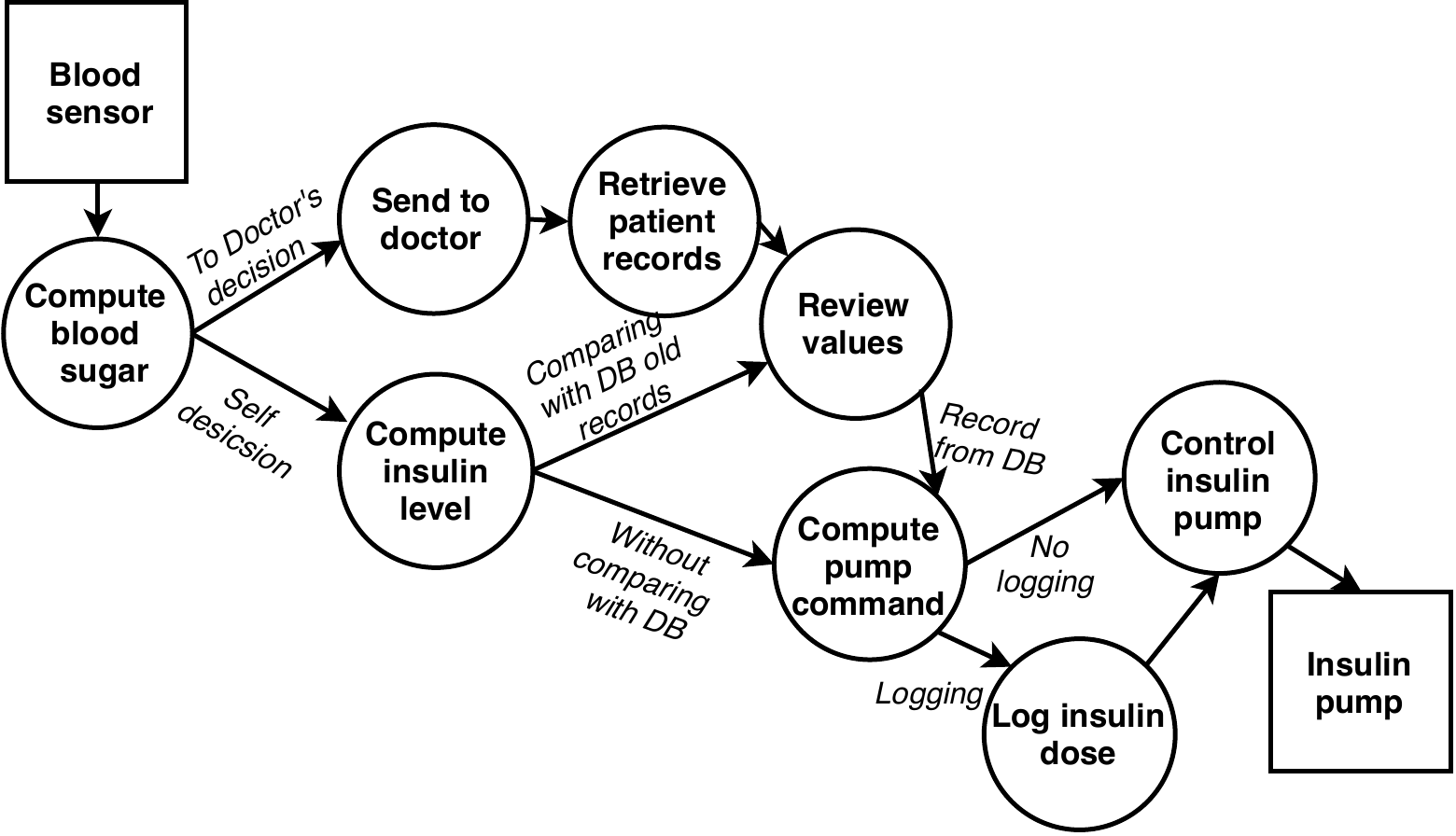}
	\caption{Insulin pump application.}
    \label{figIP}
\end{figure}

\paragraph{Insulin pump} \label{IP}
People who suffer from diabetes use a software-controlled system that must deliver the correct amount of insulin according to the current level of blood sugar. We modeled this software-based insulin pump \cite{sommerville2011software} as a set of IoT micro-sensors embedded in the patient's body that measure blood parameters proportional to the sugar level. In the Fog analytic layer, the system learns the patient state variation, computes the proper level of insulin upon abnormal state detection and sends it to the pump controller through eight components orchestrated as in Figure~\ref{figIP}.

\begin{figure}[t]
	\centering
    \includegraphics[width=0.8\columnwidth]{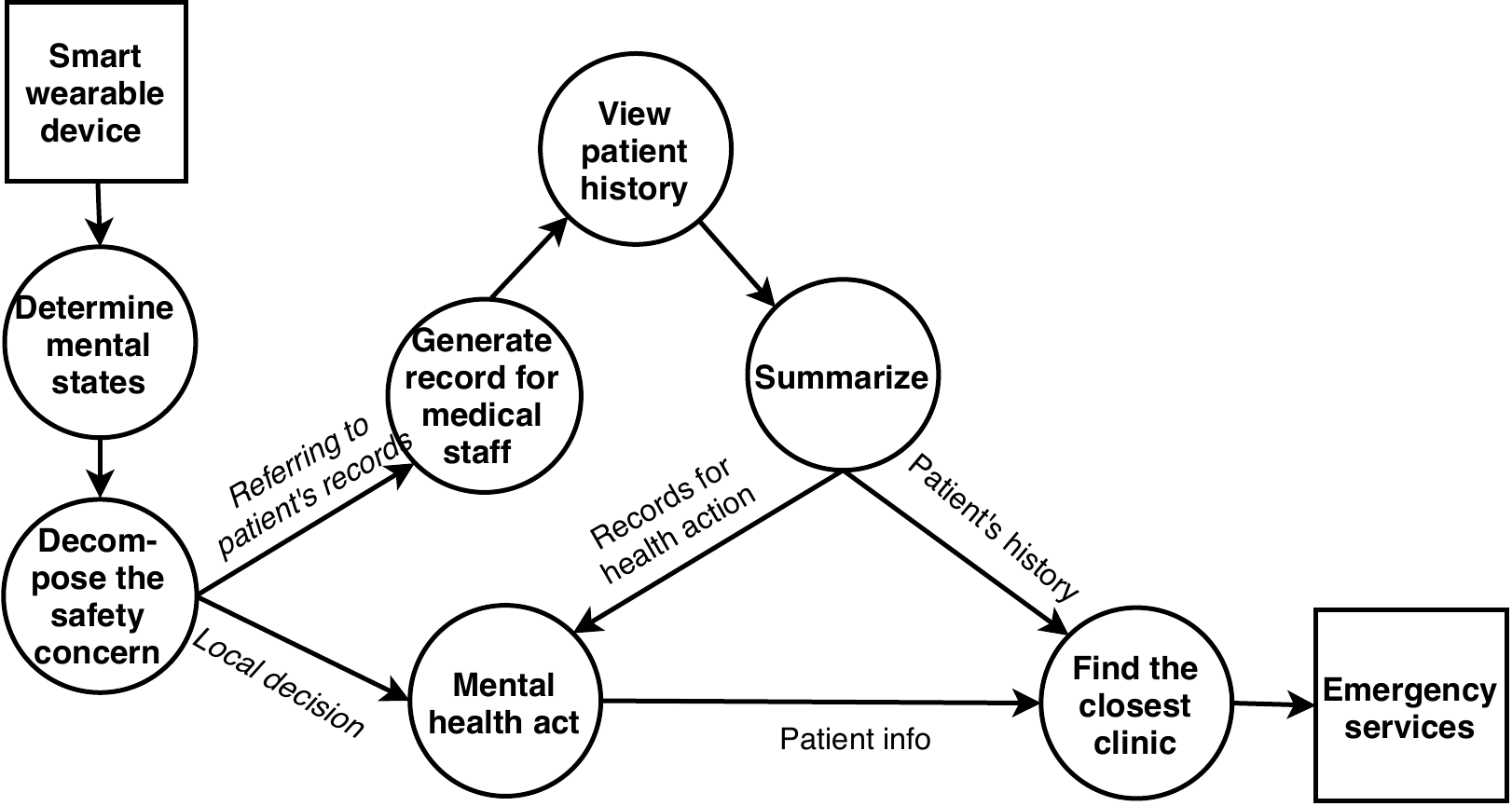}
	\caption{IoT based mental health care application.}
    \label{figMHC}
\end{figure}

\paragraph{Mental health care} \label{MHC}
This application \cite{sommerville2011software} is used in a number of UK hospitals\footnote{\url{http://iansommerville.com/software-engineering-book/case-studies/the-mentcare-system/}} for managing patient information suffering from a mental health disorder.
As such patients may not always want to attend the same clinic, they must be helped and supported through prearranged appointments, and emergency services orchestrated in a Fog environment as illustrated in Figure~\ref{figMHC}.

\section{Experimental simulation} \label{simulation}
We implemented MAPO in the jMetal multi-objective optimization framework \cite{Durillo2011JMetal} and
created elaborate scenarios based on a simulated Fog environment\footnote{https://github.com/vindem/sleipnir}, which extends the work in \cite{Brogi2017}. 
\subsection{Experimental design} \label{experimntdesign}
We investigate the benefits of MAPO for application placement compared to two state-of-the-art methods: Fog Service Placement Problem (FSPP) \cite{Skarlat2017QoS-aware} based on linear integer programming and Edge-ward (EW) \cite{gupta2017ifogsim} that implements a hierarchical best fit algorithm. We consider that EW places the last component of every application on the ISP GW because of its higher computational resources.
We evaluate MAPO considering the application completion time, energy consumption and economic cost required for executing an application, starting from the moment an IoT device or end-user provided the input data to the IoT application until another device or user collected the final output.

We designed two sets of experiments according to the characteristics of the low latency IoT applications described in \cite{Skarlat2017QoS-aware}.
The first experimental set investigates the impact to the objectives of the data size $Data_i$ transferred between the application components set to $\{0.5, 1, 4\}\ \SI{}{\mega\bit}$, with a fixed application CPU workload of $INSTR\left(m_i\right)=2000$ MI.
The second set of experiments evaluates the impact of the CPU workload $INSTR\left(m_i\right) \in \{250, 500, 1000, 2000\}$ MI by bounding the data size $\Data_i = \SI{4}{\mega\bit}$. The objective results are then averaged over $1000$ runs for statistical significance. 

\subsection{Simulator setup}

Table \ref{tbl:infra} and Table \ref{tbl:commun} display the computation and communication capabilities of the Fog devices divided in three hierarchical categories based on their computing and storage capabilities: ISP gateways (ISP GW), local WiFi gateways (WiFi GW) or cellular base transceiver station (BTS) and Mobile Edge (ME) devices.
For the Cloud and ISP GW devices, we use a configuration equivalent to a system with Intel$^\circledR$ Xeon family (i.e. Xeon Platinum 8175).
The WiFi GWs are based on an Intel$^\circledR$ Core$^{(TM)}$ i7-8550U CPU equivalent configuration. The mobile edge devices are represented either as Raspberry Pi (RPi) single-board computers or mobile phones based on ARM Cortex-A75 architectures with Qualcomm$^\circledR$ Kryo$^{(TM)}$ 385 equivalent cores. 
The devices are connected through Ethernet, Wireless LAN, or 4G/LTE network interfaces.

\begin{table}[t]
\small
\makegapedcells
\centering
\caption{Fog infrastructure configuration.}
\label{tbl:infra}
\begin{tabular}{|@{}c@{}|@{}c@{}|@{}c@{}|@{}c@{}|@{}c@{}|}
\cline{2-5}
\multicolumn{1}{{c|}}{}&{\emph{Cloud}} & \multicolumn{3}{{c|}}{\emph{Fog}}\\
\hline
\emph{Characteristic}&{CDC} &\makecell{ISP GW}&\makecell{WiFi GW/\\Cellular BTS}&{\makecell{ME}}\\
\hline
\makecell{\emph{CPU} {[MIPS]{$\cdot 10^{3}$}}} & 250 & 65 & [10,15] & [2,10]\\
\hline
\makecell{\emph{RAM} \SI{}{[\giga\byte]}} & 32 & 16 & [8,16] & [0.5,2]\\
\hline
\makecell{\emph{Storage} \SI{}{[\giga\byte]}} & 512 & 250 & 128 & 16-64\\
\hline
\makecell{$\varrho_{j}^p$ \SI{}{[\watt]}} & 1650& 530& [380,410] & [2.50,3.20]\\
\hline
\makecell{{$CP_j$ [\cent]}} & 0.03 & 0.035 & \makecell{[0.04, 0.05]} & \makecell{[0.02,0.04]}\\
\hline
{\makecell{$CS_j$[\cent]}} & 10E-7 & 15E-6 & \makecell{[10E-6,20E-6]} & \makecell{[20E-6, 30E-6]}\\
\hline
\end{tabular}
\end{table}\par

\begin{table}[t]
\small
\makegapedcells
\centering
\caption{Fog communication configuration.}
\label{tbl:commun}
\begin{tabular}{|@{}c@{}|@{}c@{}|@{}c@{}|@{}c@{}|@{}c@{}|@{}c@{}|}
\cline{2-6}
\multicolumn{1}{{c|}}{}& {\emph{Cloud}} & \multicolumn{4}{{c|}}{\emph{Fog}}\\
\hline
\emph{Characteristic} & CDC & \makecell{ISP GW}&\makecell{WiFi GW/\\Cellular BTS}&\multicolumn{2}{{c|}}{\makecell{ME}}\\
\hline
\emph{Connectivity} & {Wired} & {Wired}  & WiFi & WiFi & \makecell{Cellular}\\
\hline
\emph{Standard} &\makecell{IEEE\\802.3a/v}&\makecell{IEEE\\802.3a/b}&\makecell{IEEE\\802.11a/c}& \makecell{IEEE\\802.11a/c/n}&\makecell{4G/LTE}\\
\hline
\emph{BW\SI{}{[\mega\bit\per\second]}} & \makecell{10000}&\makecell{[1000,2000]}&\makecell{[400,1000]}& \multicolumn{2}{{c|}}{[250, 400]}\\
\hline
$\varrho_{j}^{m}$ \SI{}{[\watt]} & 1300 & 410& \makecell{[1.80,2.00]} & \multicolumn{2}{{c|}}{[1.00,1.50]}\\
\hline
$CR_j$ [\cent] & 3E-6 & 35E-7 & \makecell{[3E-6,5E-6]} & \multicolumn{2}{{c|}}{\makecell{[3E-6, 5E-6]}}\\
\hline
\end{tabular}
\end{table}
\footnotetext{system-on-a-chip}

We simulated the Fog by considering a single geographically bounded cluster \cite{kimovski2018adaptive} with eleven ME devices connected to 90 cameras or biomedical sensors and 90 actuators.
The Fog cluster is connected to a Cloud back-end infrastructure through a local proxy server.
We assume that the end-users and the IoT devices are close to the MEs and experience a \SI{1}{\milli\second} latency, while the latency between every ME and WiFi GW is \SI{10}{\milli\second}. We set the latencies between the WiFi GW and ISP GW to \SI{50}{\milli\second}, and between the ISP GW and CDC to \SI{100}{\milli\second}, obtained using the Global Ping Statistics in WonderNetwork\footnote{\url{https://wondernetwork.com/pings}\label{ping}}.

\subsection{Simulation results}
We analyze the MAPO results based on three criteria: communication data size, component CPU workload, and multi-objective algorithm quality and scalability.

\subsubsection{Data size}
Table~\ref{AR:DataSize}, Table~\ref{IP:DataSize} and Table~\ref{MHC:DataSize} demonstrate that the data size has marginal effect on the objective values
for the augmented reality, insulin pump and mental health care applications.

More concretely, MAPO reduces the completion time by up to 70\% compared to FSPP and EW, since they tend to place the application components on the ISP and WiFi GWs that are farther away from the IoT devices. This is not proper for deadline constrained applications such as the augmented reality, and results in higher communication latency. Unlike the related methods, MAPO searches for tradeoff placements by considering devices with low communication latency to the IoT layer and high computational speed and thus improves the total application completion time by reducing the communication latency and the components' completion time locally close to the IoT devices. 
Lastly, the results show that the data size does not have significant effect on the completion time as the IoT applications exchange small data with low latency. 

In terms of energy, EW consumes nearly 61\% less than MAPO. Contrarily, MAPO reduces the energy consumption by 17\% compared with FSPP. The only exception is the augmented reality application, for which MAPO provides more energy demanding solutions than FSPP. 
The energy consumption of MAPO is explained by the tradeoff between the increased computation time per application component and the reduced communication latency.
Therefore, the higher computation time introduced by the Fog devices translates into higher energy consumption for time constrained applications.  

Furthermore, MAPO reduces the economic cost by up to 77\% compared to EW and by 45\% compared to FSPP. The total cost is directly related to the cost for application computation and the cost for data communication among the executing components. The cost reductions provided by MAPO are due to the lower communication times 
which drastically reduce the completion times. \par

\begin{table}[t]
\caption{Augmented reality application completion time, energy consumption, and economic cost vs. data size.}
\centering
{
\resizebox{1\columnwidth}{!}{
\begin{tabular}{|c|c|c|c|c|c|c|c|c|c|c|}
\cline{2-10}
\multicolumn{1}{{c|}}{} & \multicolumn{3}{c|}{Time[\SI{}{\second}]} & \multicolumn{3}{c|}{Energy[\SI{}{\kilo\joule}]} & \multicolumn{3}{c|}{Cost[\cent]}\\ 
\hline
\backslashbox{\emph{Method}}{\emph{Data size}} & $0.5$ & $1$ & $4$ & $0.5$ & $1$ & $4$ & $0.5$ & $1$ & $4$ \\
\hline
\makecell{FSPP} & 0.490 & 0.491 & 0.497 & 18.6 & 18.7 & 18.9  & 0.2 & 0.21 & 0.23\\
\hline
\makecell{EW} & 0.489 & 0.490 & 0.50 &  18.4 & 18.5 & 19.3 & 0.21 & 0.21 & 0.2\\
\hline
\makecell{MAPO} & 0.163 & 0.163 & 0.166 & 34.5 & 34.8 & 49 & 0.11 & 0.12 & 0.16\\
\hline
\end{tabular}
}}
\label{AR:DataSize}
\end{table}

\begin{table}[t]
\caption{Insulin pump application completion time, energy consumption, and economic cost vs. data size.}
\centering
{
\resizebox{1\columnwidth}{!}{
\begin{tabular}{|c|c|c|c|c|c|c|c|c|c|c|}
\cline{2-10}
\multicolumn{1}{{c|}}{} & \multicolumn{3}{c|}{Time[\SI{}{\second}]} & \multicolumn{3}{c|}{Energy[\SI{}{\kilo\joule}]} & \multicolumn{3}{c|}{Cost[\cent]}\\ 
\hline
\backslashbox{\emph{Method}}{\emph{Data size}} & $0.5$ & $1$ & $4$ & $0.5$ & $1$ & $4$ & $0.5$ & $1$ & $4$ \\
\hline
\makecell{FSPP} & 1.276 & 1.279 & 1.296 & 58.5 & 58.8 & 61 & 0.3 & 0.3  &0.4\\
\hline
\makecell{EW} & 2.448 & 2.454 & 2.492 & 24 & 24 & 25 & 0.76 & 0.8 & 0.9\\
\hline
\makecell{MAPO} & 0.76 & 0.79 & 0.6 & 50 & 46  & 64 & 0.4 & 0.4 &0.2\\
\hline
\end{tabular}
}}
\label{IP:DataSize}
\end{table}

\begin{table}[t]
\caption{Mental health care application completion time, energy consumption, and economic cost vs. data size.}
\centering
{
\resizebox{1\columnwidth}{!}{
\begin{tabular}{|c|c|c|c|c|c|c|c|c|c|c|}
\cline{2-10}
\multicolumn{1}{{c|}}{} & \multicolumn{3}{c|}{Time[\SI{}{\second}]} & \multicolumn{3}{c|}{Energy[\SI{}{\kilo\joule}]} & \multicolumn{3}{c|}{Cost[\SI{}{\cent}]}\\ 
\hline
\backslashbox{\emph{Method}}{\emph{Data size}} & $0.5$ & $1$ & $4$ & $0.5$ & $1$ & $4$ & $0.5$ & $1$ & $4$ \\
\hline
\makecell{FSPP} & 1.36 & 1.37 & 1.40 & 75 & 76 & 79 & 0.3 & 0.4 & 0.5\\
\hline
\makecell{EW}  & 2.95 & 2.96 & 3.1 & 25.1 &  25.2 & 26.3 & 1 & 1.1 & 1.4\\
\hline
\makecell{MAPO}& 0.868 & 0.870& 0.879 & 62.5 & 63  & 65 & 0.4 & 0.4 & 0.4\\
\hline
\end{tabular}
}}
\label{MHC:DataSize}
\end{table}

\subsubsection{CPU workload}
Table~\ref{AR:CPU}, Table~\ref{IP:CPU} and Table~\ref{MHC:CPU} demonstrate that although all methods scale similarly in relation to CPU workloads, there are substantial performance differences.
We observe that MAPO reduces the completion time by up to 60\% compared to both FSPP and EW, especially for applications with complex patterns and higher number of components, such as the insulin pump and the mental health care case studies.
We explain the completion time improvements by the higher communication latency of the ISP and WiFi GWs extensively exploited by FSPP and EW. As observed, MAPO provides higher gains in terms of completion time for CPU workloads above 1000 MI.

Although MAPO outperforms FSPP and reduces the energy consumption by 23\% for mental health care application, it provides more energy demanding solutions than FSPP, specifically for the augmented reality application. In addition, as the energy evaluations show, EW consumes nearly 55\% less energy than MAPO. 
Whilst MAPO increases the computation time of the application components, it reduces the communication latency, therefore decreasing the completion time. In terms of the energy consumption, the higher computation time obtained by the execution through the Fog devices leads to higher energy consumption for a specific deadline constrained application.

Moreover, MAPO decreases the economic cost by up to 50\% compared with EW and 50\% with FSPP. MAPO performs better than EW in terms of economic cost, because it reduces the completion time, which translates into lower financial burden. However, FSPP incurs 22\% lower costs than MAPO due to the use of cheaper Cloud resource compared to the more expensive Fog, especially for high CPU workloads.

\begin{table}[t]
\caption{Augmented reality application completion time, energy consumption, and economic cost vs. CPU workload.}
\centering
{
\resizebox{1\columnwidth}{!}{
\begin{tabular}{|c|c|c|c|c|c|c|c|c|c|c|c|c|c|} 
\cline{2-13}
\multicolumn{1}{{c|}}{} & \multicolumn{4}{c|}{Time[\SI{}{\second}]} & \multicolumn{4}{c|}{Energy[\SI{}{\kilo\joule}]} & \multicolumn{4}{c|}{Cost[\SI{}{\cent}]} \\ 
\hline
\backslashbox{\emph{Method}}{\emph{CPU Wl}} &  $250$ & $500$ & $1000$ & $2000$ &  $250$ & $500$ & $1000$ & $2000$ &  $250$ & $500$ & $1000$ & $2000$ \\
\hline
\makecell{FSPP} &  0.05 & 0.124 & 0.246 & 0.5 & 1.9 & 4.7 & 9.4 & 18.7 & 0.02 & 0.05 & 0.10 & 0.2\\
\hline
\makecell{EW} & 0.05 & 0.123 & 0.245 & 0.5 & 2 & 4.7 & 9.3 & 18.4 & 0.03 & 0.06 & 0.11 & 0.2\\
\hline
\makecell{MAPO} &0.05 & 0.119 & 0.132 & 0.2 & 1.5 & 5 & 24 & 35  & 0.02 & 0.06 & 0.08 & 0.1\\
\hline
\end{tabular}}
}
\label{AR:CPU}
\end{table}
\begin{table}[t]
\caption{Insulin pump application completion time, energy consumption, and economic cost vs. CPU workload.}
\centering
{
\resizebox{1\columnwidth}{!}{
\begin{tabular}{|c|c|c|c|c|c|c|c|c|c|c|c|c|c|} 
\cline{2-13}
\multicolumn{1}{{c|}}{} & \multicolumn{4}{c|}{Time[\SI{}{\second}]} & \multicolumn{4}{c|}{Energy[\SI{}{\kilo\joule}]} & \multicolumn{4}{c|}{Cost[\SI{}{\cent}]} \\ 
\hline
\backslashbox{\emph{Method}}{\emph{CPU Wl}} &  $250$ & $500$ & $1000$ & $2000$ &  $250$ & $500$ & $1000$ & $2000$ &  $250$ & $500$ & $1000$ & $2000$ \\
\hline
\makecell{FSPP} & 0.3 & 0.5 & 0.8 & 1.3 & 3 & 15  & 29  & 59 & 0.1 & 0.1  &  0.2 & 0.3\\
\hline
\makecell{EW} & 0.4 & 0.6 & 1.2 & 2.5 & 6  &  6 & 12  & 24 & 0.04 &   0.2 &  0.4 & 0.8\\
\hline
\makecell{MAPO} & 0.2 & 0.4 & 0.5 & 0.8 & 1  & 12  & 23  & 50  & 0.1& 0.1 & 0.2 &0.4\\
\hline
\end{tabular}}
}
\label{IP:CPU}
\end{table}

\begin{table}[t]
\caption{Mental health care completion time, energy consumption and economic cost vs. CPU workload.}
\centering
{
\resizebox{1\columnwidth}{!}{
\begin{tabular}{|c|c|c|c|c|c|c|c|c|c|c|c|c|} 
\cline{2-13}
\multicolumn{1}{{c|}}{} & \multicolumn{4}{c|}{Time[\SI{}{\second}]} & \multicolumn{4}{c|}{Energy[\SI{}{\kilo\joule}]} & \multicolumn{4}{c|}{Cost[\SI{}{\cent}]} \\ 
\cline{1-13}
\backslashbox{\emph{Method}}{\emph{CPU Wl}} &  $250$ & $500$ & $1000$ & $2000$ &  $250$ & $500$ & $1000$ & $2000$ &  $250$ & $500$ & $1000$ & $2000$ \\
\hline
\emph{FSPP} & 0.41 & 0.6 & 0.8 & 1.4 & 8 & 19  & 37 & 75 & 0.1 & 0.1  &  0.2 & 0.3\\
\hline
\emph{EW} & 0.31 & 0.8 & 1.5 & 2.9 & 3  &  6 & 13  & 25 & 0.2 &   0.3 &  0.5 & 0.9\\
\hline
\emph{MAPO} & 0.25 & 0.5 & 0.6 & 0.9 & 1  & 10  & 32  & 63  & 0.1 & 0.2 & 0.2 &0.4\\
\hline
\end{tabular}
}
}
\label{MHC:CPU}
\end{table}

\subsubsection{Multi-objective algorithm} 
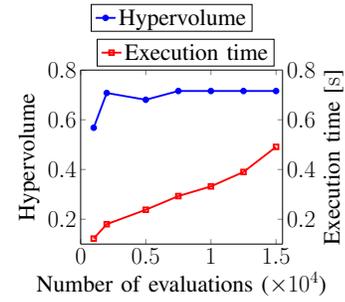
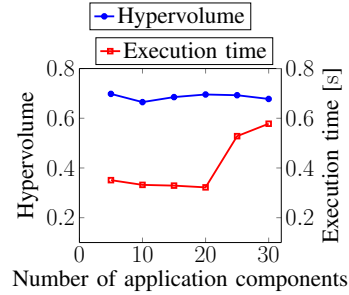
\begin{figure}
\centering
     \subfloat[\label{HV}]{
        \begin{tikzpicture}[scale=0.44]
        \begin{axis}[
        every tick label/.append style = {font=\huge},
        every axis label/.append style = {font=\huge},
        legend style={font=\huge,at={(0.05,1.3)},anchor=west},
        thick,
        scale only axis,
        xmin=0,xmax=1.55,
        ymin=0.1, ymax=0.8,
        axis y line*=left,
        xlabel=Number of evaluations ($\times 10^4$),
        ylabel=Hypervolume]
        \addplot
        [
        color=blue,
        ultra thick,
        mark=otimes*,
        ]
        coordinates{(0.1,0.567959)   (0.2,0.707828)   (0.5,0.680526)   (0.75,0.715768)   (1,0.715768) (1.25,0.715768)  (1.5,0.715768)};
        \legend{Hypervolume}
        \end{axis}
        \begin{axis}[
        legend style={font=\huge,at={(0.96,1.1)},anchor=east},
        every tick label/.append style = {font=\huge},
        every axis label/.append style = {font=\huge},
        thick,
        scale only axis,
        xmin=0,xmax=1.55,
        ymin=0.1, ymax=0.8,
        axis y line*=right,
        axis x line=none,
        ylabel={Execution time [\SI{}{\second}]}]
        \addplot[
            color=red,
            ultra thick,
            mark=square,
            ] 
            coordinates{(0.1,0.122)   (0.2,0.180)   (0.5,0.238)   (0.75,0.293)   (1,0.332) (1.25,0.390)    (1.5,0.491)};
            \legend{Execution time}
        \end{axis}
    \end{tikzpicture}
    }
    \hspace{0.000001cm}
    \subfloat[][\label{components}]{%
    \begin{tikzpicture}[scale=0.44]
    \begin{axis}[
            legend style={font=\huge,at={(0.05,1.3)},anchor=west},
            every tick label/.append style = {font=\huge},
            every axis label/.append style = {font=\huge},
            thick,
            scale only axis,
            xmin=0,xmax=32,
            ymin=0.1, ymax=0.8,
            axis y line*=left,
            xlabel=Number of application components,
            ylabel=Hypervolume]
            \addplot
            [
            color=blue,
            ultra thick,
            mark=otimes*,
            ]
            coordinates{(5,0.698033)   (10,0.664985)   (15,0.685195)   (20,0.695619)   (25,0.692716) (30,0.677695)};
            \legend{Hypervolume}
            \end{axis}
            \begin{axis}[
            legend style={font=\huge,at={(0.96,1.1)},anchor=east},
            every tick label/.append style = {font=\huge},
            every axis label/.append style = {font=\huge},
            thick,
            scale only axis,
            xmin=0,xmax=32,
            ymin=0.1, ymax=0.8,
            axis y line*=right,
            axis x line=none,
            ylabel={Execution time [\SI{}{\second}]}
            ]
            \addplot[
            color=red,
            ultra thick,
            mark=square,
            ] 
            coordinates{(5,0.351)   (10,0.332)   (15,0.329)   (20,0.322)   (25,0.528) (30,0.578)};
            \legend{Execution time}
            \end{axis}
        \end{tikzpicture}
    }
    \caption{\normalsize MAPO hypervolume and execution time of Pareto-optimal placements with different: a) number of evaluations, b) number of application components.}
    \label{MAPO}
\end{figure}

As our research deals with an NP-complete multi-objective problem, we present experiments that demonstrates the ability of MAPO to provide high quality placements across a large set of Fog devices. Figure~\ref{MAPO}a presents the correlation between the number of evaluated placements, gradually increased from 1000 to 14000 with a population of 100 individuals, the quality of the solutions quantified by the hypervolume \cite{while2006faster}, and the execution time metrics. The results show that MAPO reaches the best solutions for around 12500 evaluations, and its execution time gradually increases with the number of evaluations, although it needs only \SI{500}{\milli\second} for 14000 evaluations.
In terms of scalability for finding placements with a hypervolume of $0.7$, Figure~\ref{MAPO}b shows that MAPO's execution time increases by 43\% when the number of components increments from 5 to 30 while maintaining the quality of the solutions relatively constant.\par

\section{Real-world evaluation}\label{testbed}
To validate the simulation results, we present in this section an analysis of MAPO 
on a real-world experimental testbed.

\subsection{Experimental design}
We evaluated MAPO using the mental health care application and compared it to the related FSPP and EW approaches. 

We configured the components to generate a computational workload in the range $INSTR\left(m_i\right) \in \{250,500,
1000,2000\}$ (MI), with an input data set $Data_{i} \in \{0.5,1,4\}$ 
(\SI{}{\mega\bit}), according to the low latency IoT application characteristics described in \cite{Skarlat2017QoS-aware}. 

\begin{table}[t]
\caption{Real-world testbed configuration.}
\label{tbl5}
\small
\makegapedcells
\centering
\begin{tabular}{|@{}c@{}|@{}c@{}|@{}c@{}|@{}c@{}|}
\cline{2-4}
\multicolumn{1}{{c|}}{}& {\emph{Cloud}} & \multicolumn{2}{{c|}}{\emph{Fog}}\\
\hline
\emph{Characteristic} & CDC & \makecell{GW}&{\makecell{ME}}\\
\hline
\makecell{\emph{CPU} {[MIPS]{$\cdot 10^{3}$}}} & 250 & 65 & 65\\
\hline
\makecell{\emph{RAM} \SI{}{[\giga\byte]}} & 16 & 1 & 1\\
\hline
\makecell{\emph{Storage} \SI{}{[\giga\byte]}} & 256 & 64 & 64\\
\hline
\emph{BW\SI{}{[\mega\bit\per\second]}} &\makecell{1000}&\makecell{1000}& \makecell{1000}\\
\hline
\end{tabular}
\end{table}

\subsection{Testbed setup}
We design two sets of experiments. The first experimental set investigates the impact to the objectives of the data size $Data_i$ transferred between the application components set to $\{0.5, 1, 4\}\ 
\SI{}{\mega\bit}$, by considering a fixed application CPU workload of $INSTR\left(m_i\right)=2000$ MI.
The second set of experiments evaluates the impact of the CPU workload $INSTR\left(m_i\right) \in \{200, 500, 1000, 2000\}$ MI by bounding the data size $\Data_i = \SI{32}{\mega\bit}$.\par 
We prepared an in-lab experimental testbed consisting of four Raspberry Pi-3 B+ single-board computers (RPi), three of which act as ME and one acts as a GW.
As a CDC device, we use a virtual machine running in a private Cloud with an eight-core Intel$^\circledR$ Core$^{(TM)}$ i7-7700 CPU at \SI{3.60}{\giga\hertz} and \SI{15.6}{\giga\byte} of RAM, running Ubuntu 16.04 LTS. The testbed components were interconnected with a dedicated Gigabit Ethernet switch and secured using the SSH protocol 
(see Table \ref{tbl5}).
We assumed that the IoT devices are close to the MEs with an average latency of \SI{1}{\milli\second}. The latency between ME and GW is \SI{10}{\milli\second}, and between GW and CDC of \SI{70}{\milli\second}, obtained using the Global Ping Statistics in WonderNetwork\textsuperscript{~\ref{ping}}. For emulating the latency between devices, we created artificial network delays using the Linux \texttt{tc}\footnote{https://linux.die.net/man/8/tc} command \cite{fahs2019proximity}. We used the \texttt{nc} command for data communication between containerized application components running on different devices.\par
We installed Raspbian GNU/Linux 9.8 (stretch)\footnote{https://www.raspberrypi.org/downloads/raspbian/} 
and Docker version 18.09\footnote{https://www.docker.com/} on all RPis and deployed a containerized virtualization environment \cite{felter2015updated}.
We instantiate a Docker image on the CDC and the Fog devices by starting an Ubuntu:14.04 base image. 

\subsection{Real-world testbed results}
We examine the real-world testbed results using two parameters as in the simulation experiments: communication data size, and component CPU workload. The results from this section confirm the general trends identified in the simulation testbed.

\subsubsection{Data size}
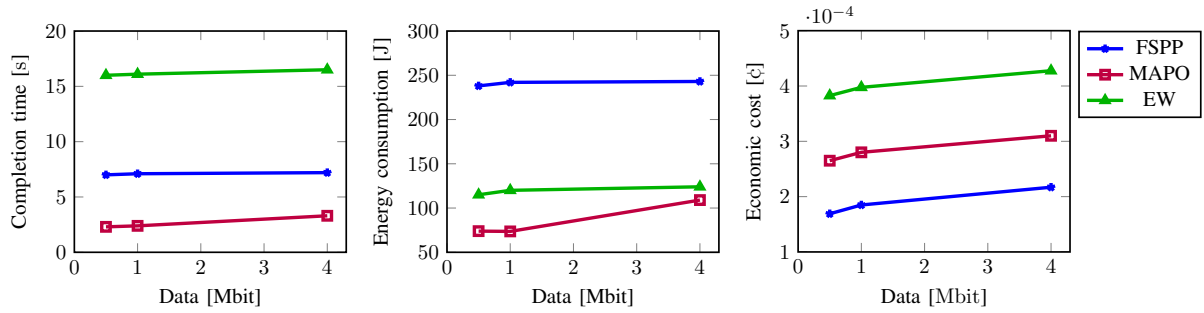
\begin{figure*}
\centering
\begin{tikzpicture}[scale=0.8]
\label{Real:Time:Data}
\begin{axis}[
    thick,
    xlabel={Data [Mbit]},
    ylabel={Completion time [\SI{}{\second}]},
    xmin=0, xmax=4.3,
    ymin=0, ymax=20,
    xtick={0,1,2,3,4},
    ytick={0,5,10,15,20},
    ]

\addplot[
    color=green!70!black,
    ultra thick,    
    mark=triangle*,
    ]
    coordinates {
    (0.5,16)(1,16.1)(4,16.5)
    };
\addplot[
    color=blue,
    ultra thick,
    mark=star,
    ]
    coordinates {
    (0.5,7.0)(1,7.1)(4,7.2)
    };
\addplot[
    color=purple,
    ultra thick,
    mark=square,
    ]
    coordinates {
    (0.5,2.3)(1,2.39)(4,3.3)
    };
\end{axis}
\end{tikzpicture}
 \hspace{0.01cm}
\begin{tikzpicture}[scale=0.8]
\label{Real:Energy:Data}
\begin{axis}[
    thick,
    xlabel={Data [Mbit]},
    ylabel={Energy consumption [\SI{}{\joule}]},
    xmin=0, xmax=4.3,
    ymin=50, ymax=300,
    xtick={0,1,2,3,4},
    ytick={50,100,150,200,250,300},
    ]

\addplot[
    color=blue,
    ultra thick,    
    mark=star,
    ]
    coordinates {
    (0.5,238)(1,242)(4,243)
    };
\addplot[
    color=green!70!black,
    ultra thick,    
    mark=triangle*,
    ]
    coordinates {
    (0.5,115)(1,120)(4,124)
    };
\addplot[
    color=purple,
    ultra thick,
    mark=square,
    ]
    coordinates {
    (0.5,73.9)(1,73.6)(4,109)
    };
\end{axis}
\end{tikzpicture}
 \hspace{0.01cm}
\begin{tikzpicture}[scale=0.8]
\label{Real:Cost:Data}
\begin{axis}[
    thick,
    legend  style={font=\normalsize,legend pos=outer north east},
    xlabel={Data [\SI{}{\mega\bit}]},
    ylabel={Economic cost [\cent]},
    xmin=0, xmax=4.3,
    ymin=0.0001, ymax=0.0005,
    xtick={0,1,2,3,4},
    ytick={0,0.0001,0.0002,0.0003,0.0004,0.0005}
    ]
\addplot[
    color=blue,
     ultra thick,
    mark=star,
    ]
    coordinates {
    (0.5,1.69e-04)(1,1.849e-04)(4,2.17e-04)
    };\addlegendentry{FSPP}
\addplot[
    color=purple,
     ultra thick,
    mark=square,
    ]
    coordinates {
    (0.5,2.64945e-04)(1,2.801e-04)(4,3.09985e-04)
    };
    \addlegendentry{MAPO}
\addplot[
    color=green!70!black,
     ultra thick,
    mark=triangle*,
    ]
    coordinates {
    (0.5,3.826e-04)(1,3.97562e-04)(4,4.2761e-04)
    };
    \addlegendentry{EW}
\end{axis}
\end{tikzpicture}
\caption{Mental health care application time, energy, and cost for different data sizes.}
\label{Real:DataSize}
\end{figure*}

Figure~\ref{Real:DataSize} shows that the completion time and the energy consumption are marginally affected by the communication data sizes, regardless of the placement method. The only exception is FSPP, which induces up to 30\% longer completion times for \SI{32}{\mega\bit} data size, as it optimizes a set of applications in a coarse-grained manner by giving priority to the CDC even for applications with small data sizes. On the other hand, there is an observable difference in the economic cost, which increases linearly with the data size.
With respect to the evaluation objectives, MAPO reduces the completion time of the application case study by up to 6.9 times compared to EW, and by 3 times compared to FSPP, which in turn performs 2.3 times better than EW. The lower completion time is due to the tendency of MAPO to optimize not only the processing time, but also the communication latency, which in turn results in a reduction of the total application completion time. In terms of the energy consumption, MAPO provides an improvement of up to 35\% compared to EW, and 68\% compared to FSPP. Finally, MAPO provides placements up to 25\% cheaper than EW, and 20\% more expensive than FSPP, which provides the most economic placements due to the frequent Cloud utilization.

\subsubsection{CPU workload}
\begin{figure*}
\centering
\begin{tikzpicture}[scale=0.8]
\label{Real:Time:CPU}
\begin{axis}[
thick,
    xlabel={INSTR [MI]},
    ylabel={Completion time [\SI{}{\second}]},
    xmin=150, xmax=2020,
    ymin=0, ymax=70,
    xtick={0,200,500,1000,2000},
    ytick={0,20,40,70},
    ]

\addplot[
    color=green!70!black,
     ultra thick,
    mark=triangle*,
    ]
    coordinates {
    (200,4.2)(500,10.2)(1000,25.1)(2000,67.5)
    };
\addplot[
    color=blue,
     ultra thick,
    mark=star,
    ]
    coordinates {
    (200,3.4)(500,6.9)(1000,15.4)(2000,40.3)
    };
\addplot[
    color=purple,
     ultra thick,
    mark=square,
    ]
    coordinates {
    (200,3.1)(500,4.4)(1000,8.2)(2000,20.1)
    };
\end{axis}
\end{tikzpicture}
\hspace{0.01cm}
\begin{tikzpicture}[scale=0.8]
\label{Real:Energy:CPU}
\begin{axis}[
    thick,
    xlabel={INSTR [MI]},
    ylabel={Energy consumption [\SI{}{\joule}]},
    xmin=150, xmax=2020,
    ymin=0.00e+3, ymax=1.550e+3,
    xtick={0,200,500,1000,2000},
    ytick={0.00e+3,0.5e+3,1e+3,1.5e+3},
    ]
\addplot[
    color=blue,
     ultra thick,
    mark=star,
    ]
    coordinates {
    (200,0.207e+3)(500,0.308e+3)(1000,0.630e+3)(2000,1.491e+3)
    };
\addplot[
    color=purple,
     ultra thick,
    mark=square,
    ]
    coordinates {
    (200,0.144e+3)(500,0.217e+3)(1000,0.462e+3)(2000,0.777e+3)
    };
\addplot[
    color=green!70!black,
     ultra thick,
    mark=triangle*,
    ]
    coordinates {
    (200,0.023e+3)(500,0.084e+3)(1000,0.202e+3) (2000,0.504e+3)
    };
\end{axis}
\end{tikzpicture}
\hspace{0.01cm}
\begin{tikzpicture}[scale=0.8]
\label{Real:Cost:CPU}
\begin{axis}[
    thick,
    legend  style={font=\normalsize,legend pos=outer north east},
    xlabel={INSTR [MI]},
    ylabel={Economic cost [\cent]},
    xmin=150, xmax=2020,
    ymin=0.0005, ymax=0.0021,
    xtick={0,200,500,1000,2000},
    ytick={0.0005,0.0010,0.0015,0.0021},
    ]
\addplot[
    color=blue,
     ultra thick,
    mark=star,
    ]
    coordinates {
    (200,5.4745e-04)(500,6.3205e-04)(1000,8.26e-04)(2000,1.366e-03)
    };
    \addlegendentry{FSPP}
\addplot[
    color=purple,
     ultra thick,
    mark=square,
    ]
    coordinates {
    (200,5.2095e-04)(500,6.0805e-04)(1000,8.0775e-04)(2000,1.3637e-03)
    };\addlegendentry{MAPO}
\addplot[
    color=green!70!black,
     ultra thick,
    mark=triangle*,
    ]
    coordinates {
    (200,6.3778e-04)(500,7.8028e-04)(1000,1.10803e-03) (2000,2.01718e-03)
    };
    \addlegendentry{EW}
\end{axis}
\end{tikzpicture}
\caption{Mental health care application time, energy, and cost for different CPU workloads.}
\label{Real:CPU}
\end{figure*}
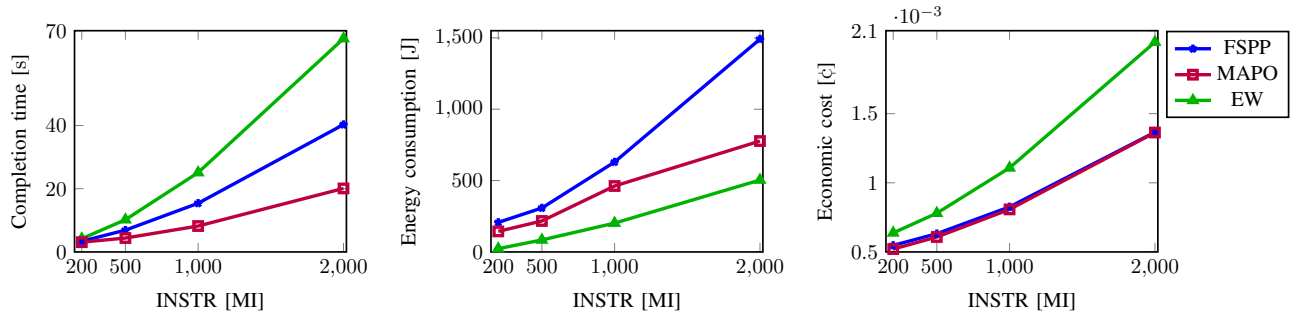

For a fixed data size of \SI{32}{\mega\bit},
Figure~\ref{Real:CPU} shows that MAPO performs two, and three times, respectively, better than FSPP and EW for varying CPU workloads of up to 2000 MI. This is due to the multi-objective nature of MAPO, which considers the component execution locality, and the communication latency together with available resources. 
Furthermore, Figure~\ref{Real:CPU} shows that MAPO provides up to 66\% higher energy efficiency than FSPP. Nevertheless, MAPO is 54\% less energy efficient than EW. For higher CPU workloads, MAPO gives preference to the energy consuming CDC device, thus achieving a tradeoff between the energy consumption and the completion time.
Finally, MAPO provides a nearly equal economic cost to the Cloud-bounded FSPP, and 27\%, on average, lower cost compared to EW as observed in the simulation.\par

\section{Conclusion and future work} \label{Conclusion}
We introduced MAPO, a multi-objective based application placement approach that considers both computation and communication aspects for executing IoT applications in a Fog infrastructure.
MAPO employs a genetic multi-objective optimization algorithm considering three conflicting criteria: completion time, energy consumption and economic cost. To solve this problem, MAPO identifies the inter-connections among the application components, searches for a Pareto set of tradeoff solutions and selects an appropriate placement. 
We evaluated MAPO for three different medical applications on a simulated and a real-world testbed infrastructure, and provided a comparison with two related methods.
Our results show that the MAPO placement is capable to reduce the application completion time 7.3 times while improving the energy efficiency, and decreasing the economic costs by up to 27\%. The results also show that the Fog infrastructure is more energy efficient for small applications that do not require high computing resources. In addition, the communication latency has a larger impact on the completion time than the communicating data size itself.

In the future, we plan to extend MAPO to support fault tolerance for application deployment in a wide area Fog environment.

\section*{Acknowledgement}
Austrian Research Promotion Agency (FFG), project 848448, Tiroler Cloud, funded this work.

\bibliographystyle{plain}
\bibliography{PlacementInFOG}

\end{document}